\documentclass[runningheads]{llncs}
\usepackage{graphicx}

\usepackage{longtable}
\usepackage{capt-of}
\usepackage{balance,bm}
\usepackage{color, colortbl, xcolor}
\definecolor{LightGray}{gray}{0.97}

\usepackage{subcaption}
\usepackage{booktabs} 
\usepackage{graphicx}
\usepackage{textcomp}
\usepackage{color,soul}
\usepackage{bm}
\usepackage{multirow}
\usepackage{wrapfig}
\usepackage{balance}
\usepackage{graphicx}  
\usepackage{enumitem}
\usepackage{wasysym}

\usepackage{colortbl}
\usepackage{arydshln}
\usepackage [english]{babel}
\usepackage [autostyle, english = american]{csquotes}
\definecolor{linkColor}{RGB}{6,125,233}
\definecolor{green}{rgb}{0.0, 0.65, 0.31}
\definecolor{bleudefrance}{rgb}{0.19, 0.55, 0.91}
\definecolor{ceruleanblue}{rgb}{0.16, 0.32, 0.75}
\definecolor{grey}{HTML}{969696}
\definecolor{violet}{HTML}{756bb1}
\definecolor{dgrey}{HTML}{01665e}
\definecolor{lgrey}{HTML}{5ab4ac}
\definecolor{dgreen}{HTML}{005a32}
\definecolor{purple}{HTML}{ae017e}

\definecolor{editCol}{HTML}{0000FF}
\definecolor{maskCol}{HTML}{c51b7d}
\definecolor{lrColor}{HTML}{8856a7}
\definecolor{trColor}{HTML}{d01c8b}
\definecolor{ctColor}{HTML}{4dac26}
\definecolor{brickred}{HTML}{f03b20}
\definecolor{improveCol}{HTML}{253494}
\definecolor{worsenCol}{HTML}{d7191c}
\definecolor{DarkBlue}{HTML}{00008B}
\definecolor{mscolor}{HTML}{01665e}
\definecolor{nmscolor}{HTML}{bf812d}
\definecolor{lgreen}{HTML}{ccece6}
\definecolor{dolive}{HTML}{308014}

\definecolor{lgreen}{HTML}{e0f3db}
\definecolor{dpink}{HTML}{CD1076}
\definecolor{pink}{HTML}{FED2D2}
\definecolor{soothinggreen}{HTML}{4dac26}
\definecolor{darkred}{HTML}{8B0000}

\definecolor{dblue}{HTML}{104E8B}
\definecolor{violet}{HTML}{8A2BE2}
\definecolor{mscolor}{HTML}{01665e}
\definecolor{nmscolor}{HTML}{d8b365}
\definecolor{deepgrey}{HTML}{525252}
\definecolor{dslate}{HTML}{2F4F4F}
\definecolor{dolive}{HTML}{556B2F}
\definecolor{teal}{HTML}{388E8E}
\definecolor{mscolor}{HTML}{01665e}
\definecolor{nmscolor}{HTML}{d8b365}

\definecolor{aicolor}{HTML}{018571}
\definecolor{occolor}{HTML}{ff7799}

\definecolor{srcolor}{HTML}{e34a33}
\definecolor{smcolor}{HTML}{253494}
\definecolor{srsmcolor}{HTML}{7fcdbb}
\definecolor{bothcolor}{HTML}{fe9929}
\definecolor{onecolor}{HTML}{018571}
\definecolor{marroon}{HTML}{881c1c}

\usepackage{array}
\usepackage{xcolor}
\usepackage{arydshln}
\usepackage{siunitx}
\colorlet{tablerowcolor4}{gray!50} 

\newcommand*{\textlabel}[2]{%
  \edef\@currentlabel{#1}
  \phantomsection
  #1\label{#2}
}
\usepackage{tcolorbox}

\colorlet{tableheadcolor}{gray!25} 
\colorlet{tablerowcolor}{gray!15} 
\colorlet{tablerowcolor2}{gray!45} 
\colorlet{tablerowcolor3}{gray!25} 

\newcommand{\rowcolmedium}{\rowcolor{tablerowcolor2}}
\newcommand{\rowcollight}{\rowcolor{LightGray}} %

\newif{\ifhidecomments}
  \hidecommentsfalse 
\ifhidecomments
    \newcommand{\shreya}[1]{}
    \newcommand{\samiha}[1]{}
    \newcommand{\dongwhi}[1]{}
    \newcommand{\koustuv}[1]{}
\else
    \newcommand{\shreya}[1]{\textbf{\small\sffamily{\textcolor{DarkBlue}{[#1 -- Shreya]}}}}
    \newcommand{\samiha}[1]{\textbf{\small\sffamily{\textcolor{dolive}{[#1 -- Samiha]}}}}
    \newcommand{\dongwhi}[1]{\textbf{\small\sffamily{\textcolor{violet}{[#1 -- Dong Whi]}}}}
    \newcommand{\koustuv}[1]{\textbf{\small\sffamily{\textcolor{dpink}{[#1 -- Koustuv]}}}}
  \fi

\colorlet{tableheadcolor}{gray!25} 
\colorlet{tablerowcolor}{gray!5} 

\definecolor{neutralCol}{HTML}{dd1c77}
\definecolor{neutralGreen}{HTML}{31a354}
\definecolor{NewBlue}{HTML}{1879ba}
\definecolor{bleudefrance}{rgb}{0.19, 0.55, 0.91}  
\definecolor{AfTrColor}{HTML}{0868ac}  
\definecolor{BfTrColor}{HTML}{a8ddb5}  

\definecolor{AfCtColor}{HTML}{b10026}  
\definecolor{BfCtColor}{HTML}{fd8d3c}

\graphicspath{ {figures/} }

\newcommand{\para}[1]{\vspace{0.5em}\noindent\textbf{#1}~}

\begin{document}
%
\title{A Checklist for Trustworthy, Safe, and User-Friendly Mental Health Chatbots}
%
%
\author{
Shreya Haran\thanks{Equal contribution.}\inst{1} \and 
Samiha Thatikonda$^{*}$\inst{1}  \and 
Dong Whi Yoo\inst{2} \and 
Koustuv Saha\inst{1}}
\titlerunning{A Checklist for Mental Health Chatbots}
\authorrunning{Haran and Thatikonda et al.}
%

\institute{University of Illinois Urbana-Champaign, Urbana, United States \and Indiana University Indianapolis, Indianapolis, United States\\
\email{sharan@illinois.edu, samihat@illinois.edu, dy22@iu.edu, ksaha2@illinois.edu}}


%
\maketitle              
%

%
\begin{abstract}


Mental health concerns are rising globally, prompting increased reliance on technology to address the demand-supply gap in mental health services. In particular, mental health chatbots are emerging as a promising solution, but these remain largely untested, raising concerns about safety and potential harms. In this paper, we dive into the literature to identify critical gaps in the design and implementation of mental health chatbots. 
We contribute an operational checklist to help guide the development and design of more trustworthy, safe, and user-friendly chatbots. The checklist serves as both a developmental framework and an auditing tool to ensure ethical and effective chatbot design. 
We discuss how this checklist is a step towards supporting more responsible design practices and support new standards for sociotechnically sound digital mental health tools.



\end{abstract}

\section{Introduction}\label{section:intro}


Mental health is a growing global concern, with conditions such as anxiety and depression affecting millions~\cite{wainberg2017challenges}. The rising demand for mental health services often outpaces the availability of qualified professionals, while the stigma surrounding mental health issues continues to impede access. These challenges emphasize the urgent need for innovative and alternative solutions~\cite{pham2022artificial}.
In recent years, mental health chatbots have emerged as tools for providing support, advice, and resources. 
The excitement about chatbots has been further amplified with the growing advancements and integrations of generative AI and large language models (LLMs) in several real-world applications~\cite{yildirim2024task,lai2023psyllm,saha2025ai,chen2020creating,yim2025generative,das2025ai}. 
For example, researchers and professionals are exploring how LLMs can be integrated into mental health care to tackle critical challenges, including resource shortages, the subjective nature of diagnoses, and the persistent stigma surrounding mental health issues. 
However, a lack of standardization and robust safeguards poses challenges in understanding and mitigating the potential harms associated with these technologies~\cite{kang2024app}.

To elaborate, chatbots are computer applications that can process and engage in a natural dialogue with an individual~\cite{mctear2016conversational}. 
A mental health chatbot can be further scoped down to a digital assistant designed to provide support, resources, and guidance for mental health illnesses. 
Prior work has shown promise at detecting specific mental health issues, teaching skills, improving confidence, etc~\cite{abd2021perceptions}. 
Chatbots can mimic human interactions in various ways, such as that of a therapist, to address people's mental health concerns~\cite{pham2022artificial,chiu2024computational,moore2025expressing,vaidyam2019chatbots}. 
These chatbots allow individuals to get immediate feedback, have a safe place to talk, spend less, and incorporate it into their daily routines~\cite{pham2022artificial,abd2021perceptions,shi2025mapping,shi2025mapping}.
However, these chatbots have faced criticism regarding their security, trustworthiness, and overall effectiveness—concerns that are especially critical in the sensitive context of mental health~\cite{khawaja2023your,moore2025expressing}. They also encounter significant challenges in handling personal data~\cite{koulouri2022chatbots,de2023benefits,yoo2025ai}.
These concerns are further complicated by the over-reliance on technology, the credibility of information, and the influence of biased or incomplete training datasets~\cite{denecke2021artificial,de2023benefits}.
Cabrera et al. identified 24 moral dilemmas of AI chatbots using bioethical principles~\cite{cabrera2023ethical}, and De Choudhury et al. analyzed the benefits and harms of large language models within an ecological framework~\cite{de2023benefits}. 
Yoo et al. explored the relationship between lived experience values, potential harms, and design recommendations for AI chatbots for mental health self-management~\cite{yoo2025ai}. 
Again, Blease and Torous
argued against using LLMs as replacements for mental health clinicians and highlighted research-identified potential harms, including algorithmic discrimination, the dissemination of false information, and breaches of privacy~\cite{blease2023chatgpt}.
Although not explicitly focused on harm prevention, 
Weisz et al. introduced design principles for generative AI applications that can be interpreted as contributing to reducing harm~\cite{weisz2024design}. 
These principles emphasize responsible design as well as shape user trust and reliance~\cite{ma2023should}.

Despite the concerns highlighted above, the development of mental health chatbots will continue, with many being integrated into existing platforms. 
Currently, there are \textit{no established standards or guidelines to regulate the development and functionality of these chatbots.}
In the absence of such regulation, these technologies may present risks and lead to negative outcomes, especially considering they serve a vulnerable population.

Prior work has proposed design principles and taxonomies for human–AI interaction~\cite{amershi2019guidelines,weisz2024design,ehsan2023charting,liao2022designing,yildirim2023investigating,das2024teacher} as well as ethical frameworks and taxonomies
for AI in mental health~\cite{coghlan2023chat,cabrera2023ethical,yoo2025ai,chancellor2019taxonomy}. 
However, these
contributions are often difficult to apply directly in practice, particularly in high-stakes mental health contexts where designers and practitioners must make concrete implementation decisions. 
Our work aims to advance this literature by translating dispersed principles into a compact, auditable tool tailored to the unique risks of mental health chatbots.


To serve as a stepping stone in bridging the above practical gaps and mitigating the potential harms of mental health chatbots, this paper proposes a checklist (Table~\ref{tab:checklist}) for the responsible design of mental health chatbots. We conducted a literature survey and identified the critical aspects and factors that chatbots need to address to function effectively and responsibly, much like a foundation for building safer, more reliable technologies.


We position this checklist as a foundational design and auditing artifact. 
By applying this checklist to an existing mental health chatbot, Woebot, demonstrates its practical applicability and reveals where existing systems struggle with subjective constructs such as transparency and empathy---highlighting the checklist's diagnostic and auditing value.
This checklist is intended to be a valuable resource for those involved in the development, use, and regulation of mental health chatbots. 
For developers and designers, it provides practical guidelines to ensure that chatbots are built with user safety, privacy, and efficacy in mind. 
End-users can also benefit by gaining a clearer understanding of what to look for in a responsible chatbot, enabling them to make informed choices about the tools they engage with. Also, regulatory bodies can use the checklist to establish standardized auditing processes that assess these chatbots' effectiveness and ethical standards. 
Therefore, this checklist can also help raise awareness and mitigate harms.

\section{Study Design and Methods}
Towards our goal of identifying guidelines and principles to help design more trustworthy, safe, and user-friendly mental health chatbots, we adopted a systematic review approach combined with thematic analysis. 
This methodology allows for a comprehensive examination of existing research, the identification of key patterns, and the development of a framework aimed at 
designing mental health chatbots.
In this section, we describe our approach below.


\subsection{Literature Review}
To conduct our literature review, we followed the Preferred Reporting Items for Systematic Reviews and Meta-Analysis (PRISMA)~\cite{moher2015preferred}---a well-established methodology to ensure rigor in selection, analysis, and reporting of the literature.
The systematic review process involved multiple key stages, which we describe below:

\para{Defining the Research Scope.} The first step in the systematic review was to define the research scope and focus on mental health chatbots. 
Specifically, we sought to analyze chatbots designed for mental health support, such as those providing interventions for anxiety, depression, stress, and general emotional well-being. We excluded chatbots with broader health purposes not focused on mental health.

\para{Literature Search.} We conducted an extensive literature search using Google Scholar as a database to identify relevant research papers. 
The search terms included ``mental health chatbots,'' ``mental health chatbots outcomes,'' ``Trustworthiness in AI,'' ``human interactions with mental health chatbots,'' ``mental health chatbots opinions of patients,'' ``mental health chatbots harms and benefits,'' ``mental health chatbots engagement and effectiveness,'' ``mental health chatbots ethics,'' and similar combinations. 
Studies published from 1989 - 2023 between a long range of time were considered, as the field of AI-driven mental health chatbots has developed significantly over the past decade. 

The inclusion criteria of our considered studies consisted of papers that---1) focus on the design, evaluation, and use of mental health chatbots, 2) address ethical issues, trust, data privacy, and the effectiveness of chatbots, and 3) discuss user experience and accessibility of mental health chatbots.
Further, the exclusion criteria consisted of papers that solely focus on general-purpose health chatbots without specific mental health interventions.



\para{Shortlisting Studies.} The above search yielded a total of 50 studies. From these, we hand-selected 43 studies for in-depth analysis given their focus on chatbot design, ethical considerations, and user experiences.
From the selected studies, we qualitatively identified key information, including---1) chatbot functionalities and interventions (e.g., CBT-based tools, emotional support, stress management), 2) reported challenges, such as data privacy concerns, user trust, and chatbot limitations, 3) metrics used to evaluate chatbots' effectiveness (e.g., user satisfaction, mental health outcomes, adherence to ethical standards),  
4) feedback from users and mental health professionals on chatbot's performance.


\para{Identifying Research Gaps.}
As part of the PRISMA-guided literature review, we systematically synthesized findings across prior studies to identify key gaps in the existing literature. This process revealed a lack of clear, standardized guidelines for ethical chatbot development, particularly with respect to trustworthiness and data security.

\para{Specifying the Research Question.}
We then refined our research question after discovering gaps in prior research---What guidelines would help develop more trustworthy, safe, and user-friendly mental health chatbots?

\para{Thematic Analysis.} Next, we conducted a thematic analysis of the reviewed literature, looking for recurring patterns and key issues in chatbot design. These themes formed the basis of our framework for ethical and effective chatbot development. 
Some of the key themes that emerged from the analysis were:


\para{\textit{Benefits}} Chatbots 
enable immediate, 24/7 responses, reducing loneliness and stigma for those uncomfortable with in-person therapy~\cite{aggarwal2023artificial,pham2022artificial,de2024chatbots}. Their availability benefits people in remote areas, reduces waitlists, and can identify symptoms in young adults~\cite{potts2021chatbots,pham2022artificial}. Chatbots are cost-effective, help reduce demand on healthcare systems, and offer non-judgmental spaces for users to share information~\cite{pham2022artificial}. They can also promote healthy behaviors like stress reduction and provide guidance for mild to moderate mental health conditions~\cite{abd2021perceptions,koulouri2022chatbots}.

\para{\textit{Harms}}
Potential harms include overreliance on chatbots, which may lead to inaccurate diagnoses or responses, risking emotional harm or legal issues. Concerns around biased training data, lack of empathy, and shallow conversations can affect user experience~\cite{au2023ai}. Issues like data privacy, the risk of hallucinations, or a chatbot misunderstanding users highlight the need for human oversight~\cite{denecke2021artificial,li2023trustworthy}. Users might also develop unhealthy attachments or experience therapeutic misconceptions where they overestimate the chatbot’s capabilities~\cite{de2024chatbots,vaidyam2019chatbots}.

\para{\textit{Ethical Concerns}}
Ethical issues arise around data privacy and the chatbot's ability to respect user autonomy. Problems include cultural biases, ensuring informed consent, and managing power dynamics between chatbots and users. Ethical frameworks emphasize principles like non-maleficence, beneficence, and justice to ensure that chatbots are transparent, avoid harm, and treat users fairly~\cite{coghlan2023chat,pham2022artificial,sweeney2021can}. The chatbot’s inability to handle crisis situations or the potential for involuntary interventions also raises concerns.

\para{\textit{Increasing User Satisfaction}}
To improve satisfaction, chatbots should engage in deeper, more human-like conversations, use proactive small talk, and create a comfortable environment. They should be intuitive and easy to use, avoiding overwhelming users with too much or irrelevant information~\cite{abd2021perceptions}. Features like guided self-help and clinically robust content, coupled with appropriate delays, humor, and rapport-building, make the experience more enjoyable~\cite{cameron2018best,li2023trustworthy}. An embodied virtual image or high-quality interface can also enhance attractiveness and engagement.

\para{\textit{Increasing User Trust}}
To build trust, chatbots must be transparent about how they function, where they get their data from, and how accurate their responses are~\cite{ahmad2023creating}. Providing clear boundaries about their capabilities, ensuring the completeness and fairness of their training data, and incorporating empathy and accountability will improve user confidence~\cite{au2023ai,li2023trustworthy,denecke2021artificial}. Trust also grows when chatbots engage in personalized conversations, take initiative, and demonstrate explainability in their actions and responses.


\subsection{Checklist Development}
Based on our literature review and thematic analysis, we developed a framework to guide the design and development of more effective, ethical, and user-centered mental health chatbots. 
We structured the framework around the key principles identified in the themes, and it aimed to capture the following facets:

\para{Ethical and User-Centered Design.}
Guidelines for ensuring user safety and privacy, including transparency around data handling and clear boundaries for chatbot interventions.

\para{Trustworthiness and Reliability.}
We use the definition of human-computer trust as ``the extent to which a user is confident in, and willing to act on the basis of the recommendations, actions, and decisions of an artificially intelligent decision aid''~\cite{madsen2000measuring}. 
Recommendations for improving chatbot training data quality, accuracy of responses, and the inclusion of human oversight where necessary.

\para{Accessibility and Inclusivity.}
Strategies for making chatbots more inclusive, with a focus on multilingual support, simple user interfaces, and diverse mental health interventions.

The above steps helped us craft an initial checklist. We then applied this checklist to an existing chatbot, Woebot, to evaluate and refine the checklist. 
Our methodological approach ensured a rigorous, evidence-based foundation for our recommendations and helped shape practical, actionable guidelines for chatbot designers, developers, and end-users. 
\section{A Checklist for Mental Health Chatbots}
\subsection{Guidelines in the Checklist}

We distilled several key principles for designing ethical mental health chatbots into a comprehensive checklist (Table \ref{tab:checklist}), which includes the following guidelines:

\para{Be Transparent.} A central finding was the importance of transparency in chatbot design. There were several concerns among users about the accuracy of the training data source, how the chatbot generated its responses, and what user data was being collected or used~\cite{au2023ai,amershi2019guidelines}. Thus, chatbots need to openly communicate how they are trained, the data they use, and how user data is collected and processed. Users should also be informed about how the chatbot generates responses, fostering trust by clarifying the decision-making processes behind its outputs~\cite{boucher2021artificially,au2023ai,amershi2019guidelines,abd2021perceptions,li2023trustworthy}. 

\para{Set Boundaries.} Overreliance or excessive attachment can lead users to believe that they should fully trust the chatbot, even though it is not yet equipped to handle that level of trust. Especially when dealing with users that could potentially have psychiatric illness~\cite{vaidyam2019chatbots}. To avoid overreliance, it is critical to define a chatbot's limitations~\cite{amershi2019guidelines,vaidyam2019chatbots}. This includes making clear what types of situations the chatbot can handle and directing users to appropriate resources or professionals when necessary, particularly in sensitive areas like mental health support. In fact, the framing and presentation of information~\cite{van2023measurements}, as well as explicitly conveying uncertainty~\cite{kim2024m} can help calibrate user trust and reliance.

\para{Provide Limitations and Accuracy.}
With chatbots, users were unsure how much they could trust the accuracy of the information provided and were unaware of the limitations in the chatbot's responses~\cite{au2023ai}. Guided by our analysis, an important recommendation is that chatbots must be clear about their accuracy and any factors that could limit the reliability of their responses~\cite{li2023trustworthy,schoenherr2023designing}. 
Transparency in decision-making is key, allowing users to understand the reasoning behind the chatbot's advice and responses.

\para{User-Friendly Design.}
Users were often overwhelmed by the interface, and they grew frustrated with the entire experience, emphasizing the importance of building a more comfortable space~\cite{abd2021perceptions}. Ease of use is vital. Chatbots should present information in a clear, concise manner and have intuitive interfaces accessible to a wide range of users~\cite{abd2021perceptions}. This ensures that people of various backgrounds and skill levels can engage with the chatbot’s functions effectively.

\para{Have Meaningful Conversations.}
Chatbots must foster meaningful, empathetic interactions, especially when handling emotional or sensitive topics~\cite{kraus2021towards,bickmore1999small,brave2005computers}. The chatbot should engage users proactively, personalize conversations, and use techniques like humor to create rapport at approriate times~\cite{kraus2021towards,baraglia2016initiative,kraus2020effects}.
It must also recognize when to offer guidance or encourage deeper reflection, while avoiding triggering content.

\para{Have Appropriate Safeguards.}
Safety emerged as a primary concern, especially regarding data privacy and handling emergency situations~\cite{khawaja2023your}. Chatbots should protect user data and avoid providing speculative or potentially harmful responses during crises. In cases of self-harm or suicide, chatbots must be trained to detect these signals and direct users to professional help immediately.

\para{Embrace Diversity and Inclusivity.}
It is essential for chatbots to be cross-culturally adaptive and reflect diverse perspectives~\cite{au2023ai,denecke2021artificial,li2023trustworthy}. Since most generative AI models are trained on data dominated by majority groups, they risk reinforcing existing stereotypes and biases. Therefore, chatbot developers must intentionally design for cultural sensitivity and inclusivity in their interactions.

\para{Build Trust.} Trust-building is essential for sustaining user engagement~\cite{au2023ai}. This involves delivering reliable and consistent information, fostering clear communication, and actively soliciting user feedback. Providing accurate responses forms the foundation of trust, enabling users to feel confident in the chatbot's credibility.
Empowering users by allowing them to control the flow of their interactions also plays a significant role in building trust~\cite{khawaja2023your,shi2025mapping,van2023measurements}. When users can choose topics and decide how to engage with the chatbot, they feel a greater sense of autonomy. Encouraging feedback and demonstrating that it is acted upon shows users that their opinions are valued, further reinforcing trust. When appropriate, using a human-like tone helps create a more relatable and comfortable interaction, encouraging users to feel at ease with the chatbot and enhancing long-term support~\cite{kraus2021towards}.

\subsection{Applying the Checklist on an Existing Chatbot: Woebot}

To understand how our checklist would apply in the real-world, we applied our checklist on a widely used mental health chatbot called Woebot. We summarize our observations below:


Woebot provides partial transparency about its training and data sources. 
Its publicly founder stated that Woebot is trained on specialized data to recognize language patterns associated with dysfunctional thoughts~\cite{darcy2024healthcare}. 
Additionally, Woebot's development involves evidence-based approaches like CBT, IPT, and DBT, crafted by trained writers in collaboration with clinical experts. However, the specifics about clinic-derived data and the influence of these practices remain unclear.
Woebot's structured conversations use natural language processing (NLP) to deliver evidence-based tools to infer behavior and feelings. Details on the specific algorithms and methodologies behind its development, as well as user feedback in updates, are insufficiently disclosed.

According to Woebot, the platform uses survey and assessment data for personalization and adheres to data privacy standards like HIPAA and GDPR. User data is treated as Protected Health Information (PHI), not sold or shared for advertising, and is stored in secure environments. Some data, including de-identified portions of conversations, is used for additional AI training. 
Although Woebot notes that annual external assessments ensure compliance with privacy standards, it remains unclear who conduct the external assessments.
Also, Woebot claims to be trained to detect concerning language through collaboration with clinical psychologists and undergo continuous monitoring and accuracy testing. 
However, it does not provide any statistics on response accuracy, validation, or details on user feedback integration. 
Finally, the website acknowledges Woebot's limitations, such as its unsuitability for crisis situations and its role as a supplemental, not replacement, mental health tool.




Overall, we found that Woebot aligned with multiple principles outlined in our checklist, underscoring the relevance and practicality of our checklist. 
This process also helped us refine our checklist to ensure it is realistic and achievable for developers.
However, we also identified significant challenges, particularly in assessing subjective dimensions such as transparency and empathy. For instance, while Woebot scored well on transparency, crucial information was often embedded deep within its resources, reducing accessibility and usability for users. 
These findings emphasize the need for further refinement in the design and evaluation of chatbot features that directly impact user trust and engagement.


\section{Discussion}\label{sec:discussion}
\subsection{Implications}


As AI chatbots such as ChatGPT are increasingly appropriated for mental health support, it becomes evident that more specialized AI technologies for mental health are on the horizon.
Early studies highlighting this appropriation trend underscore the importance of systematically understanding the potential harms of these technologies. Such a comprehensive understanding will enable us to promote safer and more effective LLM applications in mental health. Prior empirical work has shown that AI-mediated mental health support can introduce subtle but consequential harms, including emotional invalidation, misinterpretation of vulnerability, and inappropriate reassurance, even when systems are designed with benevolent intent~\cite{kang2024app,de2023benefits}.

This paper does not take a stance on the existence of mental health chatbots or evaluate their effectiveness in improving mental wellbeing. 
Rather, motivated by the need to raise awareness of the harms of AI in sensitive contexts, we propose a preliminary set of guidelines as a checklist. This is intended as a foundational starting point for responsible chatbot design rather than a comprehensive framework.
This checklist can counterbalance the growing hype around mental health  chatbots by highlighting potential harms~\cite{kim2024mindfuldiary,sharma2024facilitating,pham2022artificial}, providing a practical tool for identifying risks during design, and encouraging meaningful discussions on harm mitigation. In particular, the checklist foregrounds language-level and interactional risks that are often overlooked by high-level ethical principles, but have been shown to meaningfully shape users’ perceptions of safety, support, and trust in AI-based mental health tools.

Further, our paper-based checklist template includes sections where practitioners and designers can envision future designs for harm mitigation. This format can also be adapted for online tools, such as FigJam, where the checklist could be utilized in collaborative design workshops.
Practitioners, designers, and researchers can use this checklist to evaluate their chatbots at any stage of the design process. 
Checklist-based artifacts have been shown to support accountability and coordination across interdisciplinary teams by translating abstract ethical concerns into concrete evaluation prompts, particularly in safety-critical domains~\cite{yildirim2024task}.
An extension can be where an individual can mark each item on the checklist as ``Yes'' or ``No'' based on their evaluation
Another extension would be to include a memo section to document issues or brainstorm solutions.
In fact, developers can leverage it to design more effective chatbots, refine their existing frameworks, and ensure continuous alignment with best practices. 

The utility of the checklist's extends beyond the design phase of mental health chatbots. It can also serve as an auditing and validation tool to evaluate existing chatbots, ensuring they meet essential standards before deployment---especially in the sensitive context of mental health applications.
At the same time, the checklist can serve as a tool to raise awareness, empowering users to better understand how AI and technology interact with mental health and fostering more informed use and engagement.
Additionally, users can use the checklist as a practical tool to evaluate whether the chatbots they interact with are safe, reliable, and uphold essential standards of transparency and empathy.
By applying the checklist to an existing chatbot, we found that while some principles are easy to implement, others require more nuanced, subjective measures.
This observation aligns with prior findings that dimensions such as emotional attunement, transparency, and trust calibration are highly context-dependent and shaped by interactional framing rather than isolated system features~\cite{van2023measurements,kim2024m}.
For instance, transparency and empathy cannot always be evaluated in a binary fashion; instead, a sliding Likert scale from 1 to 5 may provide a better reflection of user experience.
Future iterations of the checklist could therefore combine binary checks with ordinal or qualitative assessments to better capture these subjective yet safety-critical dimensions of user experience~\cite{liao2022designing}.

This checklist outlines the minimal requirements that mental health chatbots must meet to uphold safety and ethical standards. However, merely adhering to these guidelines is not enough to justify the existence of a mental health chatbot or demonstrate its effectiveness. Without a deep understanding of how sensitive populations, such as individuals with mental health conditions, perceive and interact with these chatbots, mitigating potential downstream harms will remain a significant challenge. While this paper underscores critical considerations, it does not assert that meeting these standards alone guarantees positive outcomes for users. 
Additionally, further refinement and adaptation are necessary to address better the specific use cases and needs of practitioners, designers, and researchers. Future studies should explore how these groups contextualize and integrate this checklist into their work practices~\cite{yildirim2023investigating}.


\subsection{Limitations and Future Directions}
Our study has limitations, which also suggest interesting future directions. 
We acknowledge that we cannot claim that our study is a rigorous review of the literature on mental health chatbots. 
While our approach was inspired by a systematic methodology (PRISMA), 
we focused specifically on identifying key facets and dimensions for designing ethical and responsible mental health chatbots.
Our study also motivates future research involving multi-stakeholder user-studies through surveys and interviews to gather expert and end-user feedback on the relevance and usefulness of such a checklist for mental health chatbots~\cite{sogancioglu2024fairness}.
Future work should examine how practitioners and users with lived experience interpret and apply checklist items in real-world settings, building on prior evidence that guideline uptake depends heavily on organizational context and interpretive flexibility~\cite{yildirim2024task,madaio2020co}.
This feedback will be crucial in refining the guidelines further. 
We also note that our checklist---in its current form---consists of binary (yes/no) questionnaires.
However, as also noted during our evaluation on Woebot, there are grey areas---some questions would be better answered on a continuous/Likert scale.  
Future work can evaluate such measures to understand the effectiveness of such a checklist.
\section{Conclusion}
Mental health chatbots have the potential to significantly enhance access to care, improve timeliness, and deliver tailored support. However, their development must prioritize ethical considerations and foster user trust. 
This work introduces a preliminary checklist designed to guide the creation of safer, more effective, and user-centered mental health chatbots. 
The checklist encompasses a comprehensive set of guidelines spanning multiple dimensions, including transparency, boundary setting, contextual relevance, user-friendliness, meaningful conversations, safety and ethics, diversity and inclusivity, and trust-building.
Further,  we applied this checklist to an existing chatbot, Woebot, and found that while it met several dimensions of the checklist, evaluating subjective aspects such as transparency and empathy proved to be particularly challenging.
We discussed how adherence to the checklist guidelines enables developers to create tools that are not only reliable and efficient but also tailored to meet user needs and expectations, fostering a positive, safe, and trustworthy interaction experience.


{\small
\bibliography{0paperLNCS}}
\bibliographystyle{plain}

\newpage
\section{Appendix}
\setcounter{table}{0}
\setcounter{figure}{0}
\renewcommand{\thetable}{A\arabic{table}}
\renewcommand{\thefigure}{A\arabic{figure}}

\begin{longtable}{p{\columnwidth}} 
\caption{A checklist highlighting the key principles for ethical, user-friendly, and effective mental health chatbot design.}
\label{tab:checklist}\\
\toprule
\rowcolmedium \textbf{Be Transparent} \\
\rowcollight \textit{Training and Data} \\
\Square{} Be open about the training data sources.\\
\Square{} Provide an overview of the training process, including the algorithms and methodologies used.\\
\Square{} Inform users on what data is collected by the chatbot and how it will be used.\\

\rowcollight \textit{Algorithm} \\
\Square{} Describe how the chatbot processes inputs and generates responses.\\
\Square{} Explain the mechanisms of how responses are formulated and what factors influence them.\\

\rowcolmedium \textbf{Set Boundaries} \\
\rowcollight \textit{Capabilities and Limits} \\
\Square{} Avoid overreliance and misconceptions by clearly communicating the limitations of the chatbot.\\
\Square{} Be clear on the functional scope of the chatbot and explain situations the chatbot is equipped to deal with.\\
\Square{} Establish plans of action for providing users with other resources or trained professionals when necessary.\\

\rowcolmedium \textbf{Provide Context} \\
\rowcollight \textit{Accuracy and Reliability} \\
\Square{} Provide detailed information on performance metrics (e.g., accuracy), user feedback, and field-testing evaluations.\\
\Square{} Describe known limitations to the chatbot's responses and factors that contribute to these limitations.\\

\rowcollight \textit{Decision-Making} \\
\Square{} Provide sufficient details on the chatbot's decision-making process in responding to a query.\\
\Square{} Ensure that the decision-making process is transparent to users.\\

\rowcolmedium \textbf{User-Friendly Design} \\
\Square{} Provide information in a clear and concise manner to avoid overwhelming users.\\
\Square{} Design the user interface to be intuitive and accessible to users with diverse abilities and needs.\\

\rowcolmedium \textbf{Have Meaningful Conversations} \\
\rowcollight \textit{Emotional Sensitivity} \\
\Square{} Provide empathetic responses and demonstrate understanding in emotional situations.\\
\Square{} Foster meaningful discussions by guiding conversations toward reflective engagement.\\
\Square{} Exercise care to avoid emotionally triggering users during sensitive discussions.\\

\rowcollight \textit{Conversation Flow} \\
\Square{} Use appropriate delays to mimic natural conversational pacing.\\
\Square{} Adjust responses to positively influence user mood when appropriate.\\
\Square{} Proactively engage users rather than only reacting to prompts.\\
\Square{} Build rapport with users over time.\\
\Square{} Personalize responses appropriately (e.g., use of names).\\
\Square{} Ensure conversations are clinically informed, robust, and relevant.\\

\rowcolmedium \textbf{Be Safe and Ethical} \\
\rowcollight \textit{Data Privacy} \\
\Square{} Ensure user data is kept confidential and not exploited.\\

\rowcollight \textit{Emergency Situations} \\
\Square{} Detect and appropriately respond to emergency language, including suicidal ideation or self-harm.\\
\Square{} Avoid hallucination in unfamiliar situations and direct users to professional support.\\
\Square{} In high-risk situations, guide users to emergency services or crisis resources (e.g., calling 911).\\

\rowcolmedium \textbf{Embrace Diversity and Inclusivity} \\
\Square{} Use inclusive design practices that minimize bias.\\
\Square{} Incorporate diverse perspectives in training data beyond majority populations.\\
\Square{} Ensure training data supports a range of mental health conditions and lived experiences.\\

\rowcolmedium \textbf{Build Trust} \\
\Square{} Provide accurate and reliable information consistently.\\
\Square{} Maintain dependable availability and response quality.\\
\Square{} Give users control over interaction topics and engagement style.\\
\Square{} Enable feedback mechanisms and demonstrate responsiveness to user input.\\
\Square{} Reinforce progress and effort through positive feedback.\\
\Square{} Use a human-like tone when appropriate to foster comfort and trust.\\

\rowcolmedium \textbf{Other Recommendations} \\
\Square{} Design chatbots to complement---not replace---existing professional care and support resources.\\
\Square{} Enable direct connections to trained professionals when needed.\\

\bottomrule
\end{longtable}

\end{document}